\documentclass[epj,nopacs,final]{svjour}[]
\usepackage{epsfig,array}
\newcommand{\etac}{\eta_{c}}
\newcommand{\kk}{K\bar{K}}
\newcommand{\pp}{\pi\pi}
\newcommand{\jpsi}{J/\psi}
\newcommand{\EE}{e^+e^-}
\newcommand{\MM}{\mu^+\mu^-}
\newcommand{\GG}{\gamma\gamma}

\newcommand{\KK}{K^+K^-}
\newcommand{\pip}{\pi^+}
\newcommand{\pin}{\pi^-}
\newcommand{\pio}{\pi^0}

\newcommand{\jpsito}{J/\psi \rightarrow }

\newcommand{\beq}{\begin{equation}}
\newcommand{\eeq}{\end{equation}}
\newcommand{\beqn}{\begin{eqnarray}}
\newcommand{\eeqn}{\end{eqnarray}}
\newcommand{\beqns}{\begin{eqnarray*}}
\newcommand{\eeqns}{\end{eqnarray*}}
\newcommand{\bfg}{\begin{figure}}
\newcommand{\efg}{\end{figure}}
\newcommand{\bitm}{\begin{itemize}}
\newcommand{\eitm}{\end{itemize}}
\newcommand{\bnum}{\begin{enumerate}}
\newcommand{\enum}{\end{enumerate}}
\newcommand{\btbl}{\begin{table}}
\newcommand{\etbl}{\end{table}}
\newcommand{\btbu}{\begin{tabular}}
\newcommand{\etbu}{\end{tabular}}
\newcommand{\kp}{K^+}
\newcommand{\kn}{K^-}
\newcommand{\ks}{K^{0}_{S}}
\newcommand{\kstp}{K^{*+}}

\newcommand{\etap}{\eta^{\prime}}

\newcommand{\be}{\begin{enumerate}}
\newcommand{\ee}{\end{enumerate}}
\newcommand{\bi}{\begin{itemize}}
\newcommand{\ei}{\end{itemize}}

\newcommand{\ksks}{K^{0}_{S}K^{0}_{S}}

\newcommand{\uu}{\mu^+\mu^-}
\newcommand{\epen}{e^+e^-}

\date{\today}
\title{\bf \boldmath Search for $\etac$ Decays into $\pp$ and $\kk$ }
\author
{
M.~Ablikim \inst{1}               
 \and J.~Z.~Bai \inst{1}               
 \and Y.~Ban \inst{11}
 \and J.~G.~Bian \inst{1}             
 \and X.~Cai \inst{1}
 \and H.~F.~Chen \inst{16}
 \and H.~S.~Chen \inst{1}              
 \and H.~X.~Chen \inst{1} 
 \and J.~C.~Chen \inst{1}
 \and Jin~Chen  \inst{1}
% ...
}
\institute
{
Institute of High Energy Physics, Beijing 100049, People's Republic of China
\and China Center for Advanced Science and Technology(CCAST), Beijing 100080, People's Republic of China
\and Guangxi Normal University, Guilin 541004, People's Republic of China
\and Guangxi University, Nanning 530004, People's Republic of China
\and Henan Normal University, Xinxiang 453002, People's Republic of China
\and Huazhong Normal University, Wuhan 430079, People's Republic of China
\and Hunan University, Changsha 410082, People's Republic of China
\and Liaoning University, Shenyang 110036, People's Republic of China
\and Nanjing Normal University, Nanjing 210097, People's Republic of China
\and Nankai University, Tianjin 300071, People's Republic of China
\and Peking University, Beijing 100871, People's Republic of China
\and Shandong University, Jinan 250100, People's Republic of China
\and Sichuan University, Chengdu 610064, People's Republic of China
\and Tsinghua University, Beijing 100084, People's Republic of China
\and University of Hawaii, Honolulu, HI 96822, USA
\and University of Science and Technology of China, Hefei 230026, People's Republic of China
\and Wuhan University, Wuhan 430072, People's Republic of China
\and Zhejiang University, Hangzhou 310028, People's Republic of China
}
\vspace{0.2cm}

\abstract{
Using 58 million $\jpsi$ events taken with the BESII detector, a
search for $\etac$  CP violating decays into $\pp$ and $\kk$ has been
performed. No clear $\etac$ signal is observed, and upper limits for
$B(\etac\to\pp)$ and $B(\etac\to\kk)$ are given at the $90\%$
confidence level, $B(\jpsi\to\gamma\etac)\cdot B(\etac\to \pip
\pin)<1.1\times10^{-5}$, $B(\jpsi\to\gamma\etac)\cdot
B(\etac\to\pio\pio)<0.71\times 10^{-5}$, $B(\jpsi\to\gamma\etac)\cdot
B(\etac\to\kp\kn)<0.96\times 10^{-5}$ and $B(\jpsi\to\gamma\etac)\cdot
B(\etac\to K_S^{0} K_S^{0})<0.53\times10^{-5}$.
}
\titlerunning{\bf Search for $\etac$ Decays into $\pp$ and $\kk$ }
\mail{zhangbx@ihep.ac.cn or fangss@ihep.ac.cn}
\begin{document}
\maketitle
\section{Introduction}   \label{introd} 
Finding the source of CP violation is one of the most important goals
of particle physics. Violation of CP symmetry has important
consequences; it is one of the key ingredients for the
baryon-antibaryon asymmetry in our universe.  Early in 1964, CP
violation was discovered in the neutral K meson system~\cite{JHC}, and
later experimental groups provided evidence for direct CP
violation~\cite{HVA}. Almost four decades after the original
discovery, CP violation in the B meson system has been
established~\cite{BAP}.  CP violation can be experimentally searched
for in processes, such as meson decays~\cite{Kleinknecht:2003td} and
neutrino oscillation~\cite{Guglielmi:2005hi}, and by measuring the
electric dipole moments of neutrons~\cite{Harris:1999jx},
electrons~\cite{Regan:2002ta} and atoms~\cite{Romalis:2000mg}.

A total of 58 million $\jpsi$ events has been collected with the
updated Beijing Spectrometer (BESII)~\cite{besii}, and this sample
offers opportunities to search for new physics in $\jpsi$ and $\etac$
decays. In this letter, a search for CP violating $\etac$ decays into
$\pi\pi$ and $K\bar{K}$ is reported using $\jpsi \to \gamma\etac$
decays.

\section{The BES detector}  \label{BESD} 
BESII is a conventional solenoidal magnet detector that is
described in detail in Ref.~\cite{besii}. A 12-layer vertex
chamber (VC) surrounding the beam pipe provides track and trigger
information. A forty-layer main drift chamber (MDC), located
radially outside the VC, provides trajectory and energy loss
($dE/dx$) information for tracks over $85\%$ of the
total solid angle.  The momentum resolution is
$\sigma _p/p = 0.017 \sqrt{1+p^2}$ ($p$ in $\hbox{\rm GeV}/c$),
and the $dE/dx$ resolution for hadron tracks is $\sim 8\%$. 
An array of 48 scintillation counters surrounding the MDC  measures
the time-of-flight (TOF) of tracks with a resolution of
$\sim 200$ ps for hadrons.  Radially outside the TOF system is a 12
radiation length, lead-gas barrel shower counter (BSC).  This
measures the energies of electrons and photons over $\sim 80\%$ of the 
total solid angle with an energy resolution of 
$\sigma_E/E=22\%/\sqrt{E}$ ($E$ in GeV). Outside of the solenoidal coil, 
which provides a 0.4~Tesla magnetic field over the tracking volume,
is an iron flux return that is instrumented with
three double layers of  counters that
identify muons of momentum greater than 0.5~GeV/$c$.

In the analysis, a GEANT3-based Monte Carlo simulation program
(SIMBES) with detailed consideration of detector performance is
used. The consistency between data and Monte Carlo has been checked in
many high purity physics channels, and the agreement is
reasonable~\cite{NIM}. The detection efficiency for each decay mode is
determined taking into account the decay angular distributions in the
Monte Carlo simulation. The angle ($\theta$) between the directions of
$e^{+}$ and $\etac$ in the laboratory frame is generated according to a
$1+\cos^{2}\theta$ distribution, and uniform phase space
is used for the $\etac$ decaying into $\pi\pi$ and
$K\bar{K}$.

\section{\boldmath Event selection}
Charged tracks are reconstructed using hits in the MDC, and they are
required to have a good helix fit and satisfy $| \cos \theta | < 0.8$,
where $\theta$ is the polar angle of the track. For all decays except
$\jpsi\to\gamma\ksks$, the point of closest approach of each track to
the beam line must satisfy $\sqrt{x_{0}^{2}+y_{0}^{2}}<2$cm,
$|z_{0}|<20$cm, where $x_{0}$ and $y_{0}$ are the coordinates
transverse to the beam line and $z_{0}$ is the distance along the beam
line from the interaction point.

A neutral cluster is considered to be an isolated photon candidate when
the energy deposited in the BSC is greater than 0.05~GeV, the first
hit is in the beginning six radiation lengths, the angle between
the nearest charged track and the cluster is greater than
$18^{\circ}$, and the difference between the angle of the cluster
development direction in the BSC and the photon emission direction is
less than $30^{\circ}$.  More than one photon per event is allowed
because of the possibility of fake photons coming from
interactions of charged tracks with the shower counter and from
other background sources.

\subsection{\boldmath $\etac\to\pp $}
To select $\etac\to\pip\pin$ candidates, the total number of hit
layers in the muon counter is required to be less than four in order
to remove $(\gamma)\uu$ events, and the sum of
$E_{\pi^{+}}/P_{\pi^{+}}$ and $E_{\pi^{-}}/P_{\pi^{-}}$ is required to
be less than 1 to remove the large background from (radiative) Bhabha
events, where $E_{\pi^{+}}$($E_{\pi^{-}}$) and
$P_{\pi^{+}}$($P_{\pi^{-}}$) are the energy measured in the BSC and
the momentum of the $\pi^{+}$($\pi^{-}$), respectively.  At least one
track is required to be identified as a pion using TOF or $dE/dx$
information.  In order to reduce background from events with $\pio$,
$P^{2}_{tr}=4|\vec{P}_{miss}|^{2}\sin^{2}\theta_{\gamma}/2 $ is
required to be less than 0.002~(GeV/c)$^{2}$, where $\vec{P}_{miss}$
is the missing momentum of the charged particles, and
$\theta_{\gamma}$ is the angle between the missing momentum and the
photon direction.  To further suppress the dominant $\rho\pi$
background, events with more than one photon and satisfying
$|m_{\gamma\gamma}-0.135|<0.065$~GeV/$c^{2}$ are rejected, where
$m_{\gamma\gamma}$ is the invariant mass of two isolated
photons. Finally, to obtain better mass resolution and to suppress
backgrounds further, events are kinematically fitted with four
constaints (4C) to the $J/\psi \to \gamma \pi^+ \pi^-$ and
$J/\psi \to \gamma K^+ K^-$ hypotheses and are required to satisfy
$\chi^{2}_{\gamma\pip\pin} < 10$ and
$\chi^{2}_{\gamma\pip\pin}<\chi^{2}_{\gamma\KK}$. If there is more
than one photon, the fit is repeated using all permutations, and the
combination with the lowest fit $\chi^{2}$ is retained.  The
$\pip\pin$ invariant mass spectrum of events surviving selection is
shown in Fig. 1(a). No $\etac$ signal is observed.

For $\etac\to\pio\pio$, the $\pio$ mesons are
identified through $\pio\to\GG$. 
Four constraint  kinematic fits to $\jpsito\gamma\pio\pio$ are
performed using all combinations of five photons, and the combination 
of photons with the smallest $\chi^{2}$ is selected and is required
to satisfy 
$\chi^{2}_{5\gamma}<10$. To select $\pio$s, the combination with the
smallest $\delta_{\pio}$ is chosen, where  
\begin{center}
$\delta_{\pio}=\sqrt{(m_{\gamma_{1}\gamma_{2}}-m_{\pio})^{2}+
(m_{\gamma_{3}\gamma_{4}}-m_{\pio})^{2}}$,
\end{center}
and $|m_{\gamma_{1}\gamma_{2}}-m_{\pio}|<0.065$~GeV/$c^{2}$ and
$|m_{\gamma_{3}\gamma_{4}}-m_{\pio}|<0.065$~GeV/$c^{2}$ are
required. To reduce background from events with $\omega$, events with the
invariant mass of a $\pio$ and one photon satisfying
$|m_{\gamma\pio_{1(2)}}-m_{\omega}|<0.030$~GeV/$c^{2}$ are rejected.
The $\pio\pio$ mass spectrum after applying these selection criteria
is shown in Fig.1(b). No evident $\etac$ is observed in the mass
spectrum.

\begin{figure*}[htbp]
\centerline{\hbox{\psfig{file=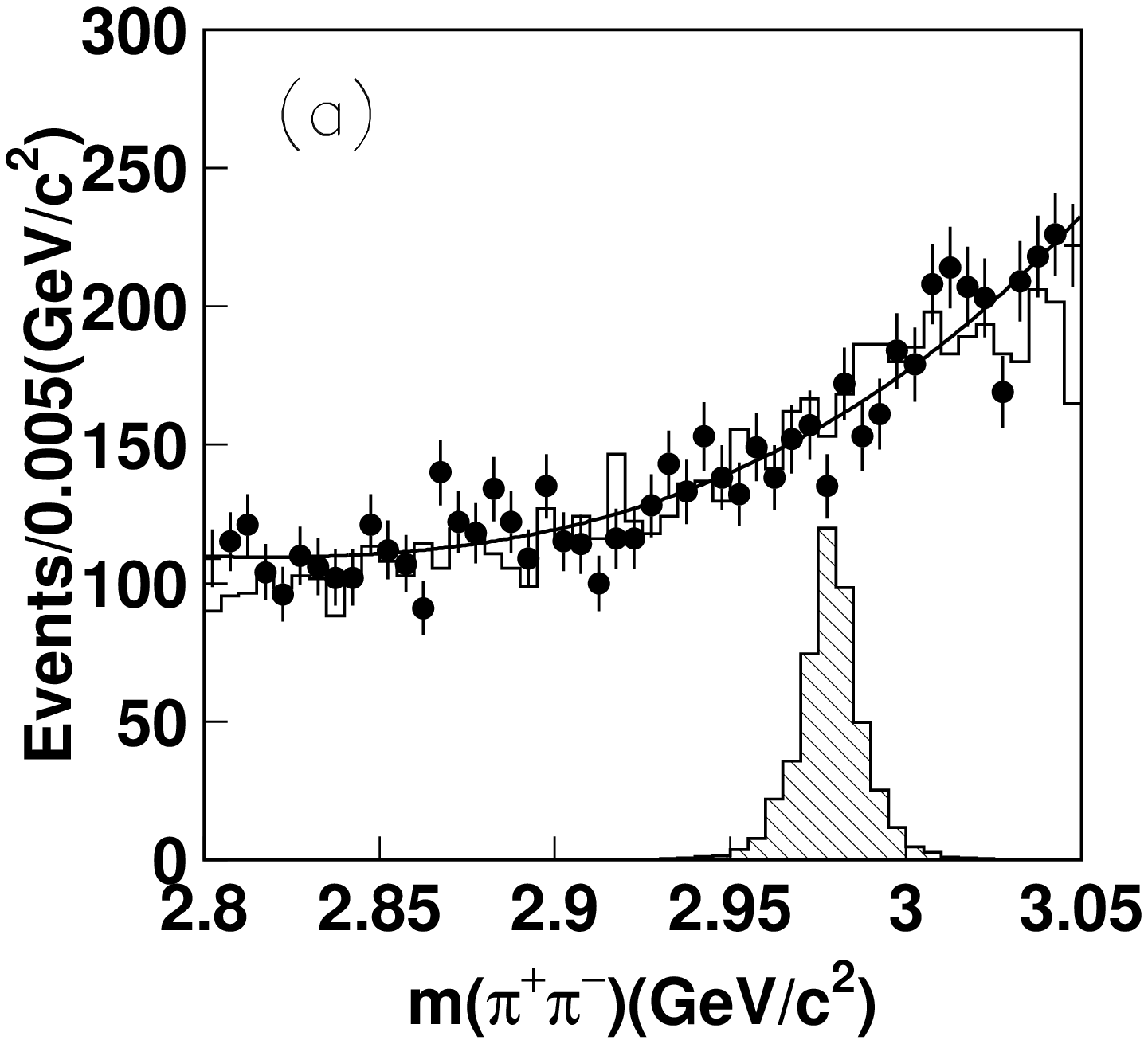,width=7.0cm}}
\hbox{\psfig{file=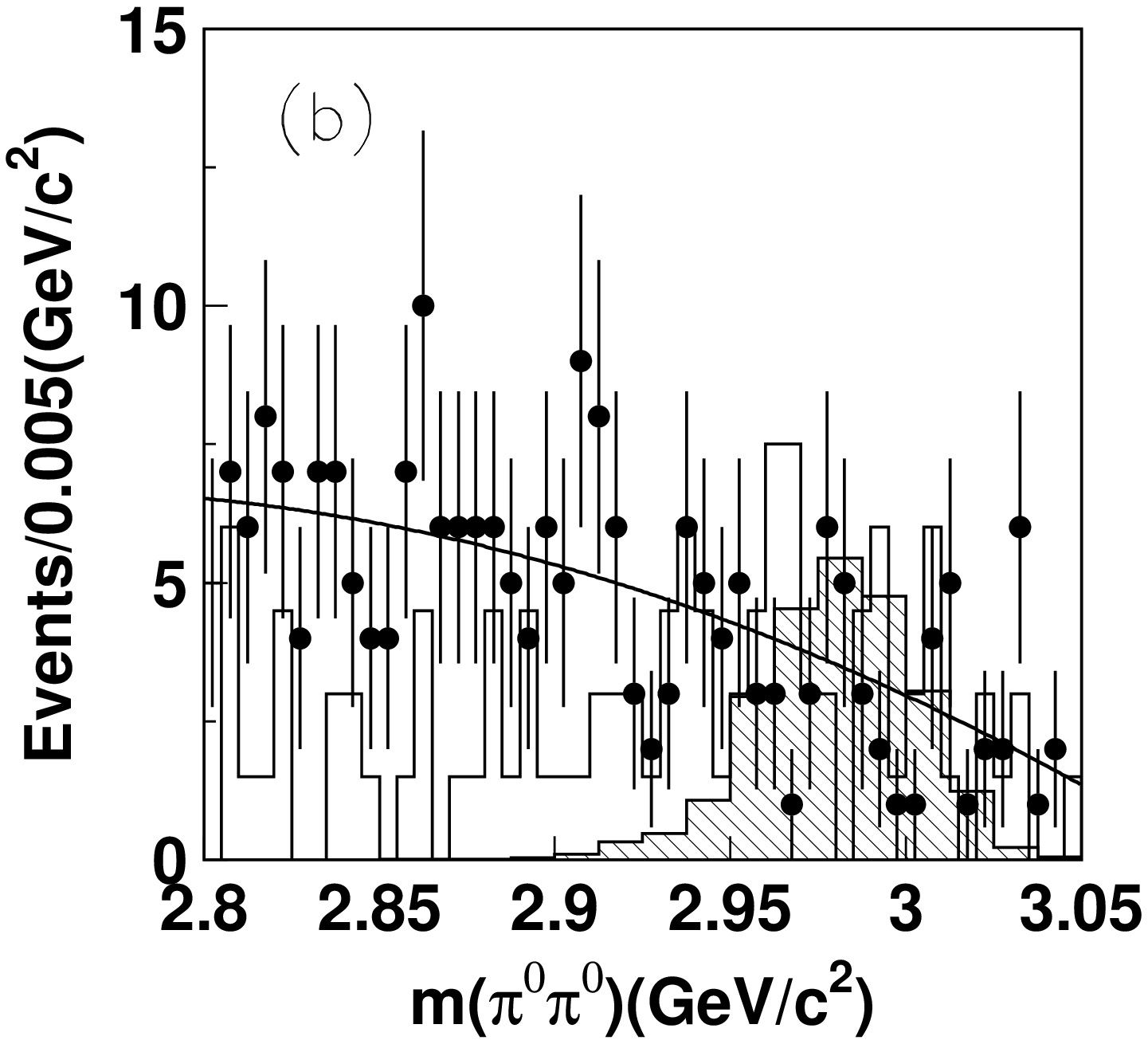,width=7.0cm}}}
\caption{Distributions of $\pi\pi$ invariant mass near the $\etac$ mass
region, where  (a) is for $\etac\to \pip\pin$ and (b) is for  
$\etac\to\pio\pio$. The points with error bars are data, and the histograms
are Monte Carlo simulated background. The smooth curves in the plots are the
best fits to the data, while the shaded histograms are the expected 
shapes of an $\etac$ signal as determined by Monte Carlo simulation (not
normalized). }
\end{figure*}

\subsection{\boldmath $\etac\to \kk $}
The total number of hit layers in the muon counter is required to be
less than four to remove $\jpsito\gamma\uu$ background.  Candidate
$\etac\to\kp\kn$ events are required to have $P^{2}_{tr} <
0.002$~(GeV/c)$^{2}$ to eliminate $\pio$ background.  To further
reduce $\jpsito\pio\KK$ and $\jpsito\pio\pip\pin$ contamination,
events surviving the above criteria and with two or more photons are
kinematically fitted to these hypotheses, and events with a fit
$\chi^{2}<50$ and with a photon pair invariant mass within
0.065~GeV/$c^{2}$ of the $\pio$ mass are rejected. Finally, to obtain
better mass resolution and to suppress backgrounds further, the two
charged tracks and one photon in the event are 4-C kinematically
fitted to the $J/\psi \to \gamma \pi^+ \pi^-$ and $J/\psi \to \gamma
K^+ K^-$ hypotheses and $\chi^{2}_{\gamma\KK}<10$ and
$\chi^{2}_{\gamma\KK}<\chi^{2}_{\gamma\pip\pin}$ are required.

The $\kp\kn$ mass spectrum is shown in Fig.2(a). 
No $\etac$ signal is evident.

For the decay $\etac\to \ks\ks$, $\ks$
mesons are identified through $\ks\to\pip\pin$. 
To ensure that the two charged pions are from the same $\ks$ vertex, 
$|z_{1}-z_{2}|$ and $|z_{3}-z_{4}|$ are required to be less than 0.06 m,
where $z_{1}$ and $z_{2}$ are the $Z$ coordinates of point of closest approach of the
track to the beam axis of the first two pions, $z_{3}$ and $z_{4}$ are 
those of the second two pions.
Candidate events must  satisfy $\delta^{2}_{\ks}<(0.020{\rm GeV}/c^{2})^{2}$,
where
$\delta^{2}_{\ks}=[m_{\pip\pin}(1)-m_{\ks}]^{2}+[m_{\pip\pin}(2)-m_{\ks}]^{2}$
and $m_{\pip\pin}$ is calculated at the $\ks$ decay vertex. The main
backgrounds from $\gamma\ks K^{\pm}\pi^{\mp}$ and $\gamma\ks\ks\pio$ events
are suppressed by requiring the 4-C kinematic fit $\chi^{2}_{\gamma4\pi}<10$.
The distribution of $K_{S}^{0}K_{S}^{0}$ invariant mass
is shown in Fig. 2(b).  No $\etac$ signal is seen.
\begin{figure*}[htbp]
\centerline{\hbox{\psfig{file=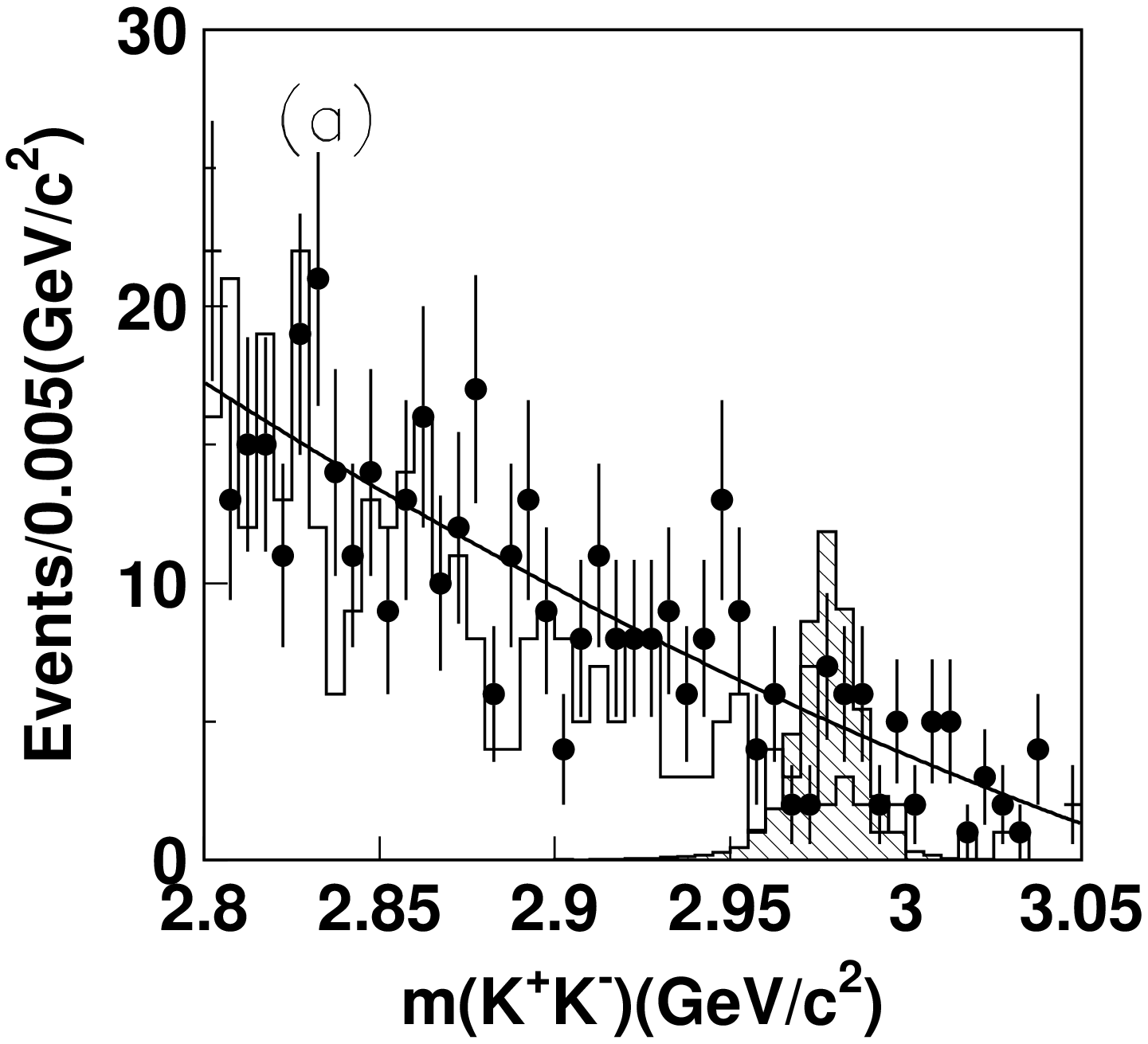,width=7.0cm}}
\hbox{\psfig{file=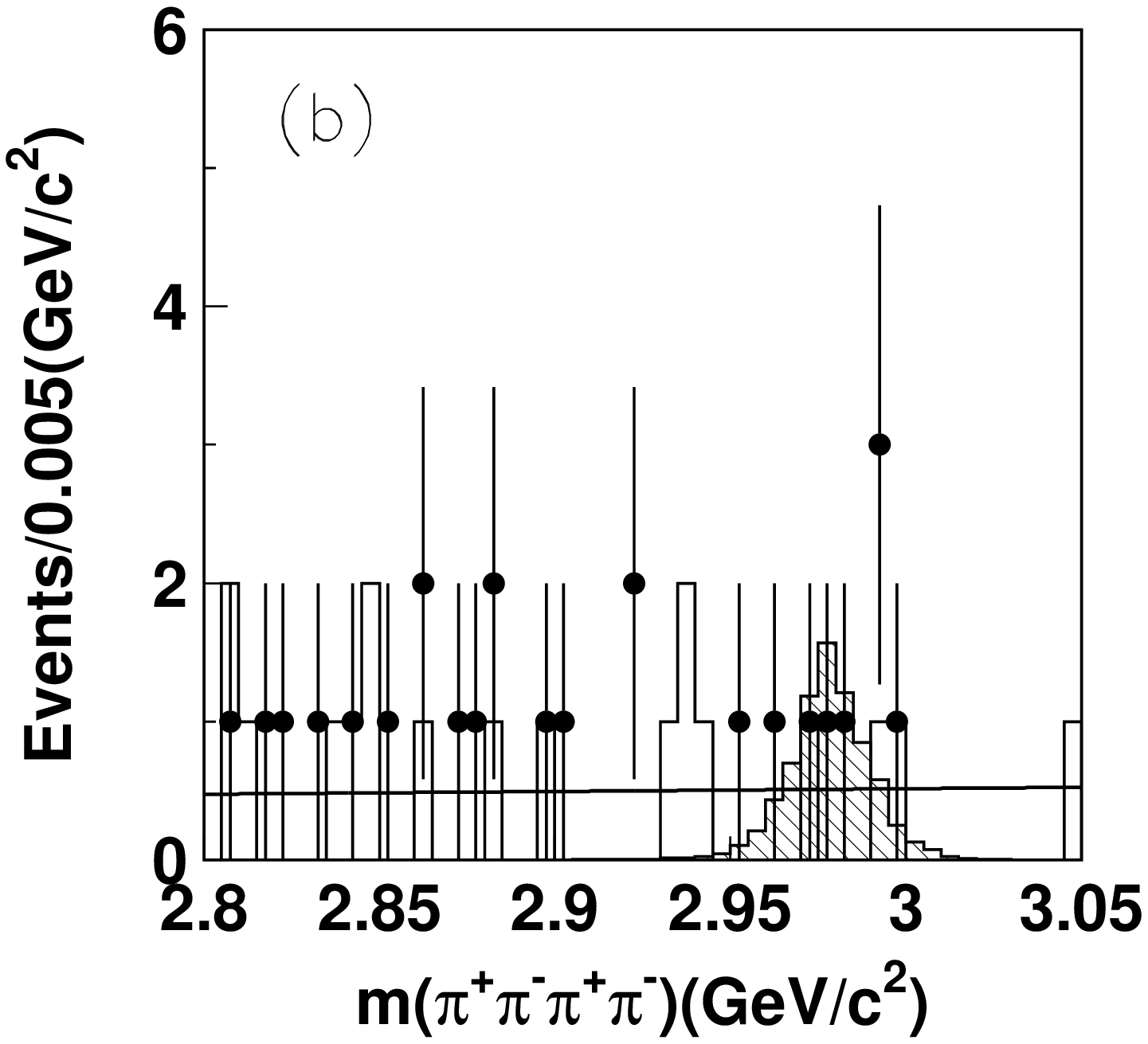,width=7.0cm}}}
\caption{Distribution of $\kk$ invariant mass near the $\etac$ mass
  region, where 
(a) is for $\etac\to\KK $ and (b) is for $\etac\to\ksks$. The points
  with error bars
are data, the histogram in (a) is Monte Carlo simulated
background, and the histogram in (b) is  background estimated using the
$\delta^{2}_{\ks}$ sideband. The smooth curves in the plots are the
best fits to the data, while the shaded histograms are the expected 
shapes of an $\etac$ signal as determined by Monte Carlo simulation (not
normalized). }
\end{figure*}

\section{\boldmath Background estimation}

Backgrounds in the various processes except $\etac\to \ksks$ are estimated 
using Monte Carlo simulation. Possible backgrounds and their contribution are 
listed in Table 1. The number of background event from Monte Carlo are normalized 
to data according to the corresponding luminosity (or known branching fraction) 
and efficiency of the samples. The main backgrounds come from $\jpsi\to(\gamma)\MM$ 
and $\jpsi\to(\gamma)\EE$ for $\etac\to\pip\pin$, from $\jpsi\to\omega\pio$ 
for $\etac\to\pio\pio$, and from $\jpsi\to K^{*+}K^{-}+c.c$ for 
$\etac\to\kp\kn$. The shape of the background from Monte Carlo simulation is 
consistent with that of data.  

For the decay $\etac\to\ksks$, non $\ks$ background is estimated using
the $\delta^{2}_{\ks}$ sideband region, defined by
$0.0008<\delta^{2}_{\ks}<0.0012$ (GeV/c$^2$)$^2$. The sideband
histogram in Fig. 2(b) has 18 background events in total. The shape of the 
background estimated by sideband is consistent with that of data. 

In fitting data, the backgrounds are fitted with polynomial function.

\begin{table*}
\begin{center}            
\caption{Dominant sources of background in the $\etac\to\pip\pin$, 
$\etac\to\pio\pio$, and $\etac\to\kp\kn$. Here $N_{data}$
and $N_{bkg}$ are the numbers of events and estimated backgrounds
in the $2\sigma$ wide $\etac$ mass region, respectively(normalized).}
\begin{tabular}{|c|c|c|c|c|c|} \hline\hline
 $\etac\to\pip\pin$  & $N_{bkg}$ &$\etac\to\pio\pio$  & $N_{bkg}$ &
 $\etac\to\kp\kn$ &  $N_{bkg}$  \\\hline        
$(\gamma)\uu$  & 585 & $\omega \pio(\gamma 2\pio)$  & 53 &
$\kstp\kn+c.c.$ & 7\\
$(\gamma)\epen$ &  281 & $\omega f_{2}(1270)(\gamma 3\pio)$ & 
11 & $\rho\pi$ & 5 \\
$\rho\pi$ &  9 &$\gamma\pio\pio\eta$ & 9 & $(\gamma)\uu$
& 5 \\
$\gamma\kp\kn$  & 5 & $\gamma\etap(\gamma\pio\pio\eta)$  & 5 &
$\gamma\pip\pin$  & 1 \\
 &  & $\omega\pio\pio(\gamma 3\pio)$ & 4 &  & \\ \hline
 sum & $880\pm 36$  &  & $82\pm 8$ &  & $18\pm1$ \\ \hline
 $N_{data}$ & 872 &  & 84 & & 25 \\
\hline\hline  
\end{tabular}
\end{center}
\end{table*}

\section{\boldmath Systematic errors}
Systematic errors on the results arise from the uncertainties in the number of
$\jpsi$ events, the secondary branching fractions of the decay modes considered, the
estimation of the selection efficiency, and the determination of the background. The various contributions to the systematic
error are listed for
all decay modes in Table 2. 
\begin{table*}            
\begin{center}
\caption{Systematic errors for the four decay modes
($\%$).}
\begin{tabular}{|c|c|c|c|c|}\hline\hline
Sources  & $\etac\to\pip\pin $  & $\etac\to\pio\pio$   
& $\etac\to\kp\kn$ &  $\etac\to\ksks$  \\\hline        
MDC tracking & 4  &  - & 4  & 8 \\
Photon efficiency &  2 & 10 & 2 & 2 \\
Kinematic fit & 5  & 5  & 5 & 5 \\
$\mu$ counter simulation & 1.2 & -  & 1.2 & - \\
$\pio$ reconstruction & -  & 0.9  & - & - \\
Mass spectrum fitting & 2.9  & 4.0 & 9.4 &5.6\\
Background uncertainty & 11.8 & 5.8 & 6.5 & 8.1 \\
Total number of $\jpsi$ & 4.7  & 4.7  & 4.7 & 4.7 \\ \hline
Total  & 14.7  & 14.1  & 14.1 & 14.6 \\
\hline\hline  
\end{tabular}
\end{center}
\end{table*}

\begin{table*}
\begin{center}            
\caption{Numbers used in the calculation of upper limits of branching fraction at the 90$\%$ 
confidence level and the upper limits for all modes.}
\begin{tabular}{|c|c|c|c|c|} \hline\hline
Decay modes & $N^{UL}_{s}$ & $\epsilon$($\%$)   & systematic error(\%)& Br.\\\hline
$\jpsi\to\gamma\etac,\etac\to\pip\pin$  & $<$ 60.2 & $10.9\pm0.1$ & 14.7 &$<
 1.1 \times 10^{-5}$ \\
$\jpsi\to\gamma\etac,\etac\to\pio\pio$  & $<$ 27.0 & $7.7\pm0.1 $ & 14.1 &
$< 0.71 \times 10^{-5}$\\
$\jpsi\to\gamma\etac,\etac\to\kp\kn$ & $<$ 30.1 & $6.3\pm0.1 $& 14.1 & $< 0.96
 \times 10^{-5}$ \\
$\jpsi\to\gamma\etac,\etac\to\ksks$  & $<$ 12.5 & $10.1\pm0.1 $ & 14.6& $<
0.53 \times 10^{-5}$ \\
$\etac\to\pip\pin$  & - & - & 34.1 & $< 1.1 \times 10^{-3}$ \\
$\etac\to\pio\pio$  & - & - & 33.9 & $< 0.71 \times 10^{-3}$\\
$\etac\to\kp\kn$    & - & - & 33.9 & $< 0.96 \times 10^{-3}$ \\
$\etac\to\ksks$     & - & - & 34.1 & $< 0.53 \times 10^{-3}$\\

\hline\hline
\end{tabular}
\end{center}
\end{table*}

\begin{enumerate}

\item The MDC tracking efficiency has been measured using channels
like $\jpsi\to\Lambda \bar{\Lambda}$ and $\psi(2S) \to \pip\pin
\jpsi$, $\jpsi\to\uu$. It is found that the Monte Carlo simulation
agrees with data within $1-2\%$~\cite{NIM} for each charged
track. Therefore $4\%$ is taken as the systematic error for
$\jpsi\to\gamma\pip\pin$ and for $\jpsi\to\gamma\kp\kn$, $8\%$ for
$\jpsi\to\gamma\ksks$.

\item The systematic error on the pion identification
efficiency is found by comparing the efficiency difference between
data and Monte Carlo.  It has been studied in detail using the decay
$J/\psi\to\rho\pi$~\cite{NIM}, where it is found that the
identification efficiency for data is in good agreement with that of
Monte Carlo. For the study of $\etac\to\pip\pin$, particle
identification is only required for one pion, and its systematic error
is negligible.

\item The systematic error from the kinematic fit has been estimated
using $\jpsi\to\rho\pi$ ~\cite{TPI} and $\jpsi\to\pip\pin\pip\pin\pio$
decays where it is possible to identify events cleanly without a
kinematic fit. The results are $4.0\%$ and $4.3\%$ for the above two
decay modes. Here $5\%$ is taken as the systematic error from the
kinematic fit for all decay modes studied in this paper.

\item The photon detection efficiency has been studied using
$\jpsi\to\rho^{0}\pio$~\cite{SML}, and the
difference between data and Monte Carlo simulation is about $2\%$ for each
photon. The resulting systematic errors on the branching fractions in
this analysis
range from $2\%$ to $10\%$ depending on the decay mode.

\item The systematic error coming from $\mu$ counter simulation is estimated
by studying the behavior of two pions in the $\mu$ counter from 
the decays $\jpsi\to\rho^{+}\pin$ and $\jpsi\to\rho^{-}\pip$. 
The result indicates that the difference 
between data and Monte Carlo simulation is $1.2\%$ 
which is taken as the systematic error.

\item The systematic error arising from $\pio$ reconstruction is
estimated by studying $\jpsi\to\rho^{0}\pio$.  We find the
$m_{\gamma\gamma}$ distribution from data is in good agreement with
that from Monte Carlo. The difference is about $0.4\%$. Therefore
$0.9\%$ is regarded as the systematic error for $\jpsi\to\gamma\pio\pio$.

\item The uncertainties of the $\etac$'s mass and width are also
systematic errors in the upper limit determination. When doing the fit
to the mass spectrum, we fix the $\etac$'s mass and width to their PDG
values. Changing the $\etac$'s mass and width by one standard
deviation from the PDG values in fitting the mass spectrum gives
systematic errors from $2.9\%$ to $9.4\%$.

\item Another systematic error arises from the uncertainty of the
background shape.  Changing the order of the background polynomial and
the fitting range gives systematic errors from 5.8$\%$ to 11.8$\%$ for
the different decay modes.

\item The uncertainty on the number of $\jpsi$ events introduces a systematic
error of $4.7\%$~\cite{FSS},  common for all decay modes. 

\end{enumerate}

The systematic errors from the above sources are listed in Table 2. The total 
uncertainty is the sum in quadrature of the individual contributions.

\section{\boldmath Results}
Since no evident $\etac$ signal is observed in the four decay modes, we
determine upper limits on the branching fractions ($B$) with the following formula
\begin{eqnarray*}
B=\frac{N^{UL}_{s}}{\epsilon\cdot N_{\jpsi}},
\end{eqnarray*}
where $N^{UL}_{s}$ is the upper limit on the number of observed events
for the $\etac$ signal, $\epsilon$ is the detection efficiency
obtained from Monte Carlo simulation, $N_{\jpsi}$ is the total number
of $\jpsi$ events, $(57.7\pm2.7)\times 10^{6}$~\cite{FSS}, which is
obtained from inclusive 4-prong hadronic decays.  For the decay
$\etac\to\ksks$, this should be divided by the square of the
$K^{0}_{S}\to \pi^{+}\pi^{-}$ branching fraction, when calculating the
branching fraction of $\etac \to \ksks$.
   
Using a Breit Wigner fit to the data, the result is consistent
with no signal. We determine a Bayesian $90\%$ confidence level upper limit
on $N_{s}$ by finding the value $N^{UL}_{s}$ such that 
\begin{eqnarray*}
\frac{\int_{0}^{N^{UL}_{s}} LdN_{s}}{\int_{0}^{\infty} LdN_{s}}=0.90,
\end{eqnarray*}
where $N_{s}$ is the number of signal events, and $L$ is the value of the 
likelihood as a function of $N_{s}$.

The values of $N^{UL}_{s}$, $\epsilon$, and the upper limits on the branching
fractions for all decay modes are listed in Table 3, where $N^{UL}_{s}$ and
$B$ are given at a Bayesian $90\%$ confidence level. The systematic 
errors are included by lowering the efficiencies by one standard deviation.
When the upper limits on B($\etac\to\pip\pin$,$\pio\pio$,$K^{+}K^{-}$,
$\ks\ks$) are derived in Table 3, an additional systematic uncertainty
of $30.8\%$ from B($\jpsi\to\gamma\etac$),$(1.3\pm0.4)\%$~\cite{PDG2004}, is included in 
the total systematic errors.

\section{\boldmath Summary}
Based on 58 million $\jpsi$ events taken at BESII, a search for $\etac$ rare decays into
$\pi\pi$ and $K \bar{K}$ has been performed. No clear $\etac$ signal
is observed, and upper
limits for $\etac\to\pi\pi$ and $\etac\to\kk$ have been obtained at
the $90\%$ confidence level for the first time
\begin{center}
$B(\jpsi\to\gamma\etac)\cdot B(\etac\to\pip\pin)<1.1\times 10^{-5}$, \\
$B(\jpsi\to\gamma\etac)\cdot B(\etac\to\pio\pio)<0.71\times 10^{-5}$, \\
$B(\jpsi\to\gamma\etac)\cdot B(\etac\to\kp\kn )<0.96\times 10^{-5}$, \\
$B(\jpsi\to\gamma\etac)\cdot B(\etac\to\ksks)<0.53\times 10^{-5}$. 
\end{center}
These results provide experimental limits
for theoretical models predicting how much CP violation there may be in $\etac$ meson decays. 

\par


\acknowledgement{{\it Acknowledgements.} 

The BES collaboration thanks the staff of BEPC for their hard
efforts. This work is supported in part by the National Natural
Science Foundation of China under contracts Nos. 10491300,
10225524, 10225525, 10425523, the Chinese Academy of Sciences under
contract No. KJ 95T-03, the 100 Talents Program of CAS under
Contract Nos. U-11, U-24, U-25, and the Knowledge Innovation
Project of CAS under Contract Nos. U-602, U-34 (IHEP), the
National Natural Science Foundation of China under Contract No.
10225522 (Tsinghua University), and the Department of Energy under
Contract No.DE-FG02-04ER41291 (U Hawaii).
}

%

\end{document}